# Cell Mechanics Based Computational Classification of Red Blood Cells Via Machine Intelligence Applied to Morpho-Rheological Markers

Yan Ge, Philipp Rosendahl, Claudio Durán, Nicole Töpfner, Sara Ciucci, Jochen Guck*, and Carlo Vittorio Cannistraci*

*Abstract* — Despite fluorescent cell-labelling being widely employed in biomedical studies, some of its drawbacks are inevitable, with unsuitable fluorescent probes or probes inducing a functional change being the main limitations. Consequently, the demand for and development of label-free methodologies to classify cells is strong and its impact on precision medicine is relevant. Towards this end, high-throughput techniques for cell mechanical phenotyping have been proposed to get a multidimensional biophysical characterization of single cells. With this motivation, our goal here is to investigate the extent to which an unsupervised machine learning methodology, which is applied exclusively on morpho-rheological markers obtained by real-time deformability and fluorescence cytometry (RT-FDC), can address the difficult task of providing label-free discrimination of reticulocytes from mature red blood cells. We focused on this problem, since the characterization of reticulocytes (their percentage and cellular features) in the blood is vital in multiple human disease conditions, especially bone-marrow disorders such as anemia and leukemia. Our approach reports promising label-free results in the classification of reticulocytes from mature red blood cells, and it represents a step forward in the development of high-throughput morpho-rheological-based methodologies for the computational categorization of single cells. Besides, our methodology can be an alternative but also a complementary method to integrate with existing cell-labelling techniques.

Y. Ge, S. Ciucci, C. Durán, and C. V. Cannistraci are with Biomedical Cybernetics Group, Biotechnology Center (BIOTEC), Center for Molecular and Cellular Bioengineering (CMCB), Center for Systems Biology Dresden (CSBD), Department of Physics, Technische Universität Dresden, Tatzberg 47/49, 01307 Dresden, Germany. (*Corresponding author: kalokagathos.agon@gmail.com)

Y.Ge, P. Rosendahl and J. Guck are with Cellular Machines Group, Biotechnology Center, Center of Molecular and Cellular Bioengineering, Technische Universität Dresden, Dresden, Germany. (*Corresponding author: jochen.guck@tu-dresden.de)

N. Töpfner is with Department of Pediatrics, University Clinic Carl Gustav Carus, Technische Universität Dresden, Dresden, Germany.

C. V. Cannistraci is with Complex Network Intelligence Center, Tsinghua Laboratory of Brain and Intelligence, Tsinghua University, Beijing, China.

## I. INTRODUCTION

IN biology, a fluorescent tag is a molecule that is chemically bound to aid in the labelling and detection of a biomolecule, and therefore serves as a label or probe. Despite the great success of fluorescent labelling, some of its shortcomings are inevitable. Some probes/labels are incompatible with live cell analysis, for example, antibody labelling against histone modifications[1], or fluorescent reporters for actin are excluded from specific filament structures during filament assembly, resulting in failed signal detection[2]. Even if live cell reporters are available[3], these may have confounding effects on the cells, such as the case of inducing single-strand DNA breaks[4] or impairing chromatin organization and leading to histone dissociation[5]. Besides, in some cases, the label can affect protein functions, or can be toxic and sometimes interfere with normal biological processes[6]. Therefore, an assay that reduces the number of, or even eliminates fluorescent labels required to identify cell phenotypes, is particularly attractive.

The call for label-free assay coincides with cell mechanical characterization. Cell mechanical properties are very often related to cell state and function, thus they can serve as an intrinsic biophysical marker[7]. As a powerful tool, cell mechanics can be used to characterize cells, to monitor their mechanical behaviour and to diagnose pathological alterations[8]. Real-time deformability and fluorescence cytometry (RT-FDC) is a microfluidic high-throughput method for morpho-rheological characterization of single cells[9]. For each cell, multiple morpho-rheological parameters are recorded in real-time and then analysed on-the-fly or in a post-processing step. In addition, also fluorescence detection and even 1-D fluorescence imaging can be performed, and the information can be correlated with the label-free morpho-rheological characterization.

In this study, we investigated how to predict cell type without fluorescence labelling by using the RT-FDC data on a case study with computational approach. To be more specific, our main aims are two: 1) to investigate the problem of computational classification of mature red blood cells (mRBCs) and reticulocytes (RETs) - derived from human

blood - considering only morpho-rheological cell features. 2) to investigate the extent to which a basic unsupervised and linear approach performs (in comparison to supervised approaches) to discriminate mRBCs and reticulocytes (RETs) on the exclusive basis of morpho-rheological phenotype data obtained from RT-FDC. We focused on this classification task because the investigation of RETs (their percentage and cellular features) in the blood is an important indicator to differentiate between multiple human diseases[10]. As reticulocyte count is an important sign of erythropoietic activity, it can help e.g. to evaluate different types of anaemia, which is a deficiency in the number or quality of red blood cells. Whereas in acute bleeding or in hemolysis the reticulocyte count is increased (or stable), a low reticulocyte count can indicate dysplastic or aplastic bone marrow disorders, resulting in an impaired erythropoiesis. In addition to quantitative changes, the RETs can change their mechanical properties and become progressively more deformable as they mature towards their normal state, a characteristic that facilitates their release from the functional healthy bone marrow[11].

Mature human red blood cells are characterized by the lack of a nucleus and consequently the absence of transcriptional activity, so that neither DNA nor RNA is typically present in these cells. In contrast, immature red blood cells can be identified by the presence of remaining amounts of nucleic acid, which can be labeled and detected using intercalating dyes such as Hoechst, DAPI or syto13. Indeed, staining of RNA in reticulocytes is a (gold-)standard procedure in clinical blood counts. Here, Nucleic acid dye, syto 13, is used as a fluorescent probe for the ground-truth label information to evaluate our classification performance. We controlled factors associated with fluorescence label issues in order to generate a bona-fide dataset. These data were obtained with a high level of confidence and low noise because the fluorescence labels were adopted according to standard procedures which ensure the respect of staining ability. In our presented pipeline, we adopted a robust unsupervised machine learning procedure and used the PC-corr[12] algorithm to extract the most discriminative markers and their correlations, which were used subsequently to classify mRBCs and RETs. Since the number of RETs in the blood is much smaller than the number of mRBCs, this classification task represents a challenging benchmark to test the proposed machine learning procedure. In addition, label-free classification of mRBCs and RETs based on cell morpho-rheological markers is a very complicated task, and as far as we know there is not any literature on the application of machine learning to this problem, therefore this represents also an innovative topic to consider for precision medicine. We successfully infer a robust combinatorial-marker (a single composed-marker that is defined as mathematical combination of several morpho-rheological markers) and define an appropriate marker threshold that can offer two-group-classification (mRBCs or RETs) of uncategorized cells with acceptable accuracy. The workflow presented hereafter can be generalized and applied to identify other cellular phenotypes (e.g., healthy vs cancer cell, marker positive vs marker negative cell) starting from multidimensional cell-mechanical measures.

## II. METHODOLOGY

### A. Ethical Statement

With ethical approval for the study (EK89032013) from the ethics committee of the Technische Universität Dresden, we obtained blood from healthy donors with their informed consent in accordance with the guidelines of good practice and the Declaration of Helsinki. However, they are regarded as three potential patients for research purpose who will go for blood check in this study. Indeed, any patient who needs a diagnosis can be healthy or pathological.

### B. Data collection and generation of training and validation set

Capillary blood was collected after finger prick from three donors (P1, P2, P3) with a 21G, 1.8 mm safety-lancet (Sarstedt AG & Co.). A volume of 2 µl blood was diluted in 1mL of 0.5% methyl cellulose complemented with 2.5 µM syto13 nucleic acid stain (Thermo Fisher Scientific Inc., S7575) and incubated 5 minutes at room temperature. RETs contain some RNA in the cytosol that they completely lose during maturation towards mRBCs, therefore they can be distinguished by RNA content. RNA staining enables measurement of RNA content which is related to the maturity of the red blood cells since they lose RNA gradually over a time of ca. one day[13]. Afterwards, all samples were measured by RT-FDC, which not only detects the mechanical phenotype of each individual cell (normal RT-DC[14], [15], characterized by ten features: area, area ratio, aspect, brightness, brightness standard deviation, deformation, inertia ratio, inertia ratio raw, x-size and y-size; see next section for more details about their descriptions) but also simultaneously gather its fluorescence intensity in a manner similar to flow cytometry. This directly correlates mechanical data with fluorescence data based on nucleic acid staining. There is a natural unbalanced cell-group composition in each donor, i.e., the percentage of RETs is much smaller than mRBCs. Since sample P1 contained more RETs in comparison to P2 and P3, we decided to adopt it for deriving the training set. Therefore, considering P1 donor, which comprises of 15,763 mRBCs and 357 RETs, 10,763 mRBCs and 257 RETs were used to create the training set named P1-partition1. The remaining 5,000 mRBCs and 100 RETs were used to create the independent *internal* (we use the word *internal* because the validation is based on cells coming from the same donor used for training) validation set, named P1-partition2. The other two donors, P2 and P3, were taken as independent *external* (because the cells are derived from donors different from the one adopted for training) validation sets, which contains 16,671 mRBCs & 145 RETs, and 15,511 mRBCs & 103 RETs, respectively. To facilitate replication of the results, these data are available for open access as supplementary data.

*C. Descriptions of the ten morpho-rheological features*

RT-DC detects the morpho-rheological properties of each single cell, which goes through the microfluidic channel, and represents them with ten numerical features:
1. area: the cell's cross-sectional area derived from the contour.
2. area ratio: the ratio between the area of the convex hull of the cell's contour and the area of the cell's contour
3. x-size: the maximal axial-length of the cell suspended in the flow that pass through the channel along the horizontal dimension
4. y-size: the maximal axial-length of the cell suspended in the flow that pass through the channel along the vertical dimension
5. aspect ratio: x-size/y-size
6. brightness: the average brightness value of the pixels inside cell's contour
7. brightness standard deviation: standard deviation of the pixels' brightness values inside cell's contour
8. deformation: the deformation of a cell is defined as D = 1 − c, where c is the circularity of the contour. Circularity is defined as: $c = \frac{2\sqrt{\pi\ area}}{perimeter}$
9. inertia ratio: ratio of the image moments[16] of the convex hull of the contour. This parameter is similar to the aspect ratio but has sub-pixel accuracy.
10. inertia ratio raw: same as above but for the raw contour (no convex hull applied)

*D. Unsupervised dimension reduction machine learning procedure*

We adopted PCA, which is a machine learning method for unsupervised linear and parameter-free dimension reduction. We performed unsupervised analysis instead of a supervised one, because it is less prone to overfitting as shown in previous studies[12], [17], [18]. 10,000 resampled datasets were generated from the original training set (P1-Partition 1), each of which was obtained by randomly selecting 200 mRBCs and 200 RETs. We will refer here and in the remainder of the text to this as class-balance procedure. PCA was used to project the data into the first three dimensions of embedding (the first three principal components). We considered only the first three dimensions of embedding since, in general, they form a reduced 3D space of representation to the original multidimensional data, where the patterns associated with the major data variability are compressed. This procedure was repeated 10,000 times, one for each of the resampled datasets. We created balanced datasets because PCA is a data-driven approach and the unbalanced datasets would impair its performance since learning algorithms often fail to generalize inductive rules over the sample space when presented with this form of imbalance[19]. We stress that the procedure to project the data is based on unsupervised dimensional reduction learning because we never used the labels to learn the multivariate transformation that projects the data onto the low-dimensional space.

Next, we considered the labels of the samples (without performing any learning procedure) just to reveal the extent to which the PC1, PC2 and PC3 are able to discriminate the two sample classes. For this, we used the p-value obtained by Mann–Whitney $U$ test[20] and AUC-ROC to evaluate the mRBCs vs RETs separation on each single dimension, and then summarized the mean p-value and the mean AUC-ROC by considering the 10,000 resampled datasets (Table I).

*E. PC-corr discriminative networks and combinatorial marker design*

PC-corr[12] is an algorithm able to enlighten discriminative network functional modules associated with the most discriminant dimension of PCA, which in our case was PC2. We applied the PC-corr algorithm to each of the 10,000 datasets and we considered the mean discriminative networks (obtained as mean of 10,000 networks) associated with the PC2 separation. We applied a cut-off of 0.6 (see Results section C. for more detail) on the weights of this mean discriminative network to extract the modules of PC2-related-features and we detected a unique discriminative network module composed by three morpho-rheological-features (Fig.2A). Then, the features that are engaged in the module of highest association with the PCA discrimination can be mathematically combined (using their mean) to offer a unique value that is named the combinatorial marker. As clarified in the result section, we considered all the possible combinations of the three morpho-rheological-features in order to design potential combinatorial markers to test in the validations. Hence, we designed four candidate combinatorial markers based on the three scaled (using z-score transformation) individual features: the mean of area, y-size and x-size; the mean of area and y-size; the mean of area and x-size; the mean of y-size and x-size.

*F. Validation of the designed combinatorial markers*

The validation set P1-partition2, which is composed of 5,000 mRBCs and 100 RETs, was used to create 10,000 resampled datasets. We randomly selected 100 samples from the 5000 mRBCs and merged them with the unique 100 RETs for each resampling population. We used the p-value obtained by Mann–Whitney U test and AUC-ROC to evaluate the classification performance of the combinatorial markers and the single markers. The mean p-value and mean AUC-ROC were calculated based on the 10,000 resampled datasets (Table II).

We clarify that the learning of the combinatorial feature selection on the P1-partion1 dataset is data-driven and unsupervised by performing the PCA and the discriminative network analysis by PC-corr. However, in the next section we describe how to supervisedly detect the optimal operator point (marker threshold) of this marker for the further class prediction on the validation datasets. Therefore, the word unsupervised in the remainder of the article refers to the way we build the marker and not to the way we select the marker threshold.

*G. Marker threshold learning and evaluation*

We used the P1-partition1 as the training set to get the optimal operator point (which is the point on the ROC curve that offers the highest AUC for the classification of the two different categories of cells) for the combinatorial markers obtained as the mean of area and y-size. In this case we used a supervised procedure (which is a hypothesis-driven procedure that exploits the training set labels to learn a threshold) and therefore we had to employ a 10-fold cross-validation (first divide the training set to ten partitions, use the nine partitions to learn the optimal operator point, and test it on the remaining one partition to get its performance according to AUC-ROC). This 10-fold cross-validation procedure was repeated ten times. Each time the ten partitions were created independently at random starting from the original training set (including 10,763 mRBCs and 257 RETs), but in this case, we preserved the original count ratio of mRBCs and RETs (10,763/257=41.9) in every partition. This means that we did not balance the training dataset by considering the same amount of mRBCs and RETs, because the procedure of learning the marker threshold is supervised and hypothesis-driven and we wanted that the optimal operator point (the marker threshold used to decide whether a cell belongs to mRBC or RET) could be learned considering the natural cell unbalance occurring in the blood samples. As a second option, we implemented the same cross-validation procedure above, but for each step we applied a class-balance procedure, such as we did for the unsupervised dimension reduction projection with mRBC resampling. This means that each cross-validation fold was composed of 257 mRBCs (sampled uniformly at random from the 10,763) and 257 RETs. Unfortunately, class-balance learning offered poor results (data not shown). This bad performance of the class-balance procedure is motivated by the fact that here the learning of a pure threshold for a marker value, and not a model to create the marker itself, is implemented. Therefore, the original cell-ratio offers advantages to learn the marker threshold value.

We obtained an array with 100 values, with each element specifying the optimal operator point generated by the respective iteration of the cross-validation procedure. By taking the median of the 100 results, which turned out to be more robust in comparison with taking average due to outliers, we estimated the overall optimal operator point value, which was considered as the most appropriate marker threshold. We used the learned marker threshold to predict the fluorescence classification in the validations of P1-partition2, P2 and P3 in two ways, and we evaluated the effectiveness of the combinatorial marker and its threshold in prediction using accuracy, sensitivity, specificity, AUC-ROC and precision. In the first way, we used these datasets by considering the natural unbalanced composition of mRBCs and RETs. More precisely, we firstly z-score-scaled each dataset and then classified each cell by comparing the learned threshold with its mean of z-score-scaled area and the z-score-scaled y-size, and computed the performance using the five abovementioned performance measures (Table III) and compared with the supervised machine learning methods described in the next section (Table IV). In the second way, we adopted the 10,000 times resampling by each time randomly taking the same amount of mRBCs with RETs in the investigated dataset (100 for P1-partition2, 145 for P2 and 103 for P3), and computed the final performance by taking the average precision (Figure 3) of the obtained 10,000 results. Also in this case comparison with the supervised machine learning methods was provided.

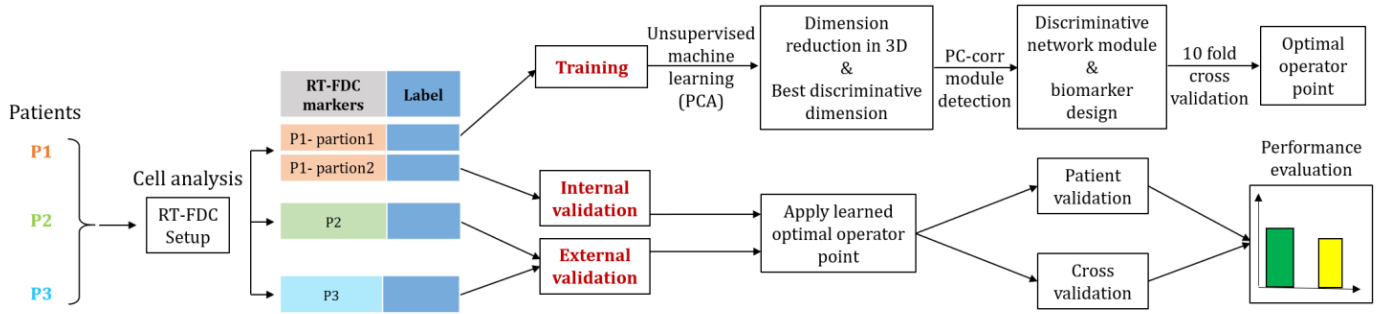

**Figure 1: Study workflow.** Blood samples were taken from three potential patients and measured using RT-FDC. The output in this case was ten morpho-rheological features together with classification information of each single cell resulting from the fluorescence signal. P1-partion1 was used for training purpose, while P1-partion2, P2 and P3 were used for validation.

*H. Other supervised machine learning methods*

The procedure to obtain the results for the supervised analysis was implemented as follows. First of all, a selection of the most important features for the segregation between classes was carried out by means of a machine learning strategy called feature selection. As we did for the learning of the marker threshold in our proposed method (see section G. above), also here we preserved the original count ratio of mRBCs and RETs (10,763/257=41.9) in every cross-validation fold. However, considering that here we learn an entire model and not only a threshold value, we obtained poor results (data not shown) and therefore we moved to adopt a class-balance procedure. In practice, for each machine learning, we trained 10 models.

Each model was trained with 10-fold cross-validation, and each fold was composed of 257 mRBCs (sampled uniformly at random from the 10,763) and 257 RETs. Then, the average model (obtained by averaging internally the parameter settings of each machine learning) of each machine learning was considered for prediction. Elastic net is a well-known algorithm that can be used for this purpose[21]. It needs a parameter called alpha that combines the L1 and L2 penalties of lasso and ridge regularization methods at different proportions. The alpha value was automatically tuned by changing its value from 0.1 until 0.9 in steps of 0.1 and the one that gave the highest AUC-ROC performance between the two classes (RETs and mRBCs), that is 0.5, was used as a parameter in Elastic net. On the other hand, Gini index [22] is a criterion used as feature selection for random forest (RF) and helps to determine which features are the most important to split the classes of the dataset, by giving them a score depending on how many trees of the random forest they were selected as a split criterion. Another feature selection strategy, and used in this case for Support vector Machine (SVM), is called recursive feature elimination (RFE). It works with the help of an external estimator, in this case SVM, that assigns weights to the features to recursively prune them until a desire number of features is eventually reached. The last feature selection strategy is intrinsically used for partial least square discriminant analysis (PLSDA) and was carried out by calculating the regression coefficients of partial least squares (PLS) and ranking them according to the number of latent variants for PLS.

In order to reduce overfitting in the feature selection, all feature selection algorithms were carried out ten times in a 10-fold cross validation (CV) procedure (a total of 100 iterations). The feature selection consists of two steps. The first step is to compute the final number (of selected features) which is fixed to the average number (that we call *m*) selected for each CV step. The second step is to determine the *m* final features to select. This is implemented by assigning to each feature an average weight obtained as the average across the weights gained in the CV steps, and then by selecting the *m* features with the highest average weights in the CV steps. Specifically, elastic net selected 7 features (area, aspect ratio, brightness, brightness SD, deformation, inertia ratio and y-size), while Gini index selected 5 (area, area ratio, brightness, brightness SD and inertia ratio), as well as RFE (area, area ratio, deformation, inertia ratio and y-size) and PLSDA (aspect, inertia ratio, inertia ratio raw, x-size and y-size).

Once the predictors (features) were chosen, the machine learning models were created in a 10-fold CV step. SVM models (features selected from elastic net and RFE) were produced with the auto optimization of hyperparameters, and with linear and non-linear (RBF) kernels. The RF model (features selected from Gini index) contains five hundred decision trees and was generated with the default parameters (fraction of input data to sample with replacement: 1; minimum number of observations per tree leaf: 1; number of variables to select at random for each decision split: 3 [that is approximatively the square root of the number of variables, which in this case is 10]) as well as PLSDA (default parameter is only the tolerant of convergence: 1E-10) and Logistic Regression (features selection method: elastic net; default parameters are the Model – we used the nominal model - and the Link function - we used logit function).

## III. RESULTS

### A. Study workflow

The overall study workflow is represented in Figure 1. The goal of this study is to investigate the ability to classify mature red blood cells (mRBCs) and reticulocytes (RETs) present in the blood of an individual (a patient who needs a diagnosis and could be healthy or pathological), considering only morpho-rheological cell features for the prediction. Hence, we emphasize that the fluorescence probe is used only for testing the performance of the prediction. The first step in the study workflow was to acquire the data from RT-FDC setup (see Methods), which can be used to analyse the presence and prevalence of all major blood cell types, as well as their morpho-rheological features, directly in whole blood[23]. In addition, it can measure the fluorescence intensity of each single cell just as in a conventional flow cytometer. The output is a 2D data matrix, where each row represents a different single cell found in the blood and the columns report for each single cell the respective morpho-rheological values (area, x-size, y-size, etc.) and the corresponding fluorescence intensity that is used to classify cells into mRBCs or RETs. We then proceeded to the unsupervised machine learning by means of PCA using P1-partion1 as training set, with the aim to find the best discriminative dimension by evaluating the separation of the mRBCs and RETs on the first three embedded dimensions. Afterwards, we applied PC-corr algorithm based on the learned best discriminative dimension to detect the discriminative network functional modules that can be used to design the combinatorial marker (because it is a combination of single morpho-rheological markers) for the classification of the two group of cells. To learn the optimal operator point that can be later used for testing, we applied 10-fold cross validation for 10 times to find the combinatorial marker threshold by using P1-partion1. Finally, we tested the performances of our defined combinatorial maker in combination with the learned optimal operator point on three independent datasets, the internal validation dataset P1-partition2 and the external validation datasets P2 and P3 with potential patient validation and cross validation.

### B. Unsupervised dimension reduction analysis by PCA and its evaluation

Due to the natural biological unbalance of mRBCs (90% to 95% of blood cells) against RETs (0.5% to 1.4%) in the blood of healthy adult donors[24], and also to prevent dataset overfitting, we performed the unsupervised learning on the P1-partition1 dataset by using a resampling procedure, which generated from P1-partition1 a total of 10,000 new resampled datasets (see section D of Methodology for detail). The final p-value and AUC-ROC are reported in Table I, and an example PCA results from the 10,000 performed PCA is shown in Supplementary Figure 1, both of which clearly indicate that the second dimension (PC2) of PCA reveals the most significant discrimination, regardless of the measure (p-value or AUC-ROC) used to assess the two-group separation.

**TABLE I**
**Evaluation of Unsupervised Machine Learning Dimension Reduction**

| PCA dimension | mean p-value | mean AUC-ROC |
|---|---|---|
| PC2 | 2.17E-13 | 0.76 |
| PC1 | 3.67E-03 | 0.61 |
| PC3 | 2.15E-02 | 0.60 |

Therefore, PC2 weights can be used to apply the PC-corr algorithm, which is able to extract a network composed of feature modules (in this case morpho-rheological measures) related to the two groups separation. We would like to emphasize that the PC-corr algorithm is not a univariate approach that selects single features independently from each other, but instead it is able to perform a multivariate prioritization that emphasizes a cohort of feature-interactions that are most discriminative according to a PCA dimension. This cohort of discriminative feature-interactions generally tends to form one or multiple discriminative network modules that — as we will illustrate in the following section — can be used for designing combinatorial markers.

*C. PC-corr discriminative module and combinatorial marker design*

We applied the PC-corr algorithm to each of the 10,000 resampled datasets and we considered the mean discriminative network (obtained as mean of 10,000 networks) associated with the PC2 separation. We applied a cut-off of 0.6 on the weights of this mean discriminative network to extract the modules of PC2-related-features. We chose the threshold of 0.6 so that extracted features have at least Pearson Correlation of 0.6 between them, and it is the highest cut-off that ensure the node connectivity, which means, there are only unconnected singular nodes with higher cut-offs (data not shown). We detected a unique discriminative network module composed of three morpho-rheological features (Fig. 2A). The "area" is the cross-sectional area, outlined by the blue contour in Fig. 2B. The "y-size" is the maximal vertical (perpendicular to flow direction) extension of the cell suspended in the liquid passing through the channel, while "x-size" is the maximal horizontal extension of the cell (Fig. 2B). PC-corr also discloses the positive correlations between the discovered features, which are represented by red edges in the network (Fig. 2A). From the ten features available, the PC-corr algorithm helps to unveil those that we should use to design the candidate combinatorial markers. We designed four candidate combinatorial markers, considering the three PC-corr selected and scaled (using z-score transformation) features: the mean of area, y-size and x-size; the mean of area and y-size; the mean of area and x-size; and the mean of y-size and x-size. In the next section we will discuss the performance evaluation and validation of these four markers, in comparison to all the original ten morpho-rheological features.

*D. Validation of the designed combinatorial markers*

To independently evaluate the classification ability of the four combinatorial markers proposed in the previous section, we considered their performance on the P1-partition2 dataset, which had not been used for learning the markers. The measures used for the evaluation are the Mann-Whitney p-value and AUC-ROC, and the results obtained for each of the four proposed markers and for each of the ten original single features are reported in Table II. Also, in this case we considered the mean performance over 10,000 resampled datasets, containing equal number of mRBCs and RETs (please refer to the methods for details). We discovered that the combination of area and y-size as a unique marker yields the best result. As a comparison, we also calculated the performance offered by the original ten features individually. Taken together, these results prove that a combinatorial selection of the features using PC-corr can tremendously improve the design of the final candidate markers. Interestingly, PC-corr pointed out a discriminative module composed by two interactions between three features: 1) area and y-size; and 2) area and x-size. Our validation in Table II disclosed that area and y-size alone are very discriminative features (AUC: 0.73 and 0.76 respectively), whereas x-size is a poor discriminative feature (AUC: 0.52). The question remains, why PC-corr included also x-size in the discriminative module? The answer is that although singularly x-size is a poor discriminative feature, PC-corr suggests not only *discriminative associations* between features, but also mechanistic relations between features in the module. In fact, the area is by definition a function of x-size and y-size, and PC-corr successfully infers this from the data independently from the single discriminative power of each feature. This result is possible because PC-corr is a multivariate approach and offers results different from univariate analysis approaches (which test single features), as extensively discussed in the article of Ciucci *et al*. [12].

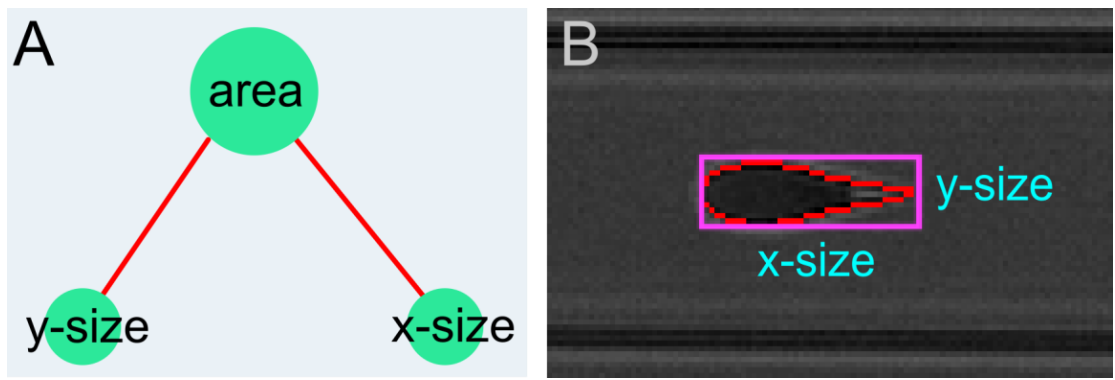

**Figure 2.** A) Discriminative network module detected by PC-corr and related with PC2 discrimination (cut-off = 0.6). B) Image of a red blood cell flowing in the RT-FDC channel, including illustrations of "area" (bounded by the blue contour), "x-size" and y-size".

**Table II**
**Classification Performance of the Candidate Markers and The Ten Single Features on The P1-partition2 Validation Dataset.** The precise meaning of each feature is given in the methods section.

| Marker | mean p-value | mean AUC-ROC |
|---|---|---|
| Average (area and y-size) | 1.47E-09 | 0.78 |
| y-size | 1.55E-08 | 0.76 |
| area | 1.54E-06 | 0.73 |
| Average (area, x-size and y-size) | 1.80E-06 | 0.73 |
| Average (x-size and y-size) | 5.71E-06 | 0.72 |
| inertia ratio | 2.69E-04 | 0.68 |
| inertia ratio raw | 3.67E-04 | 0.67 |
| aspect | 7.13E-04 | 0.66 |
| deformation | 3.70E-03 | 0.64 |
| brightness | 5.21E-03 | 0.64 |
| Average (area and x-size) | 9.10E-03 | 0.63 |
| area ratio | 1.70E-01 | 0.57 |
| x-size | 6.08E-01 | 0.52 |
| brightness standard deviation | 6.27E-01 | 0.52 |

*E. Marker threshold learning and evaluation*

Let us suppose that the morpho-rheological measures of the cell population of a new individual are provided, and that we are interested in applying the combinatorial marker based on the area and y-size (which provided the best performance in the previous evaluation) in order to classify mRBCs and RETs. Yet, what we miss is a threshold for the combinatorial marker so that we can use it to predict new unknown cell's class. In order to learn a proper marker threshold, we used again the P1-partition1 (previously adopted to learn the discriminative module) and selected as best threshold the one that corresponds to the optimal operator point (see section G of Methodology for detail). According to this procedure, we found that 0.51 was the best threshold for the designed marker, computed as the mean of the z-score-scaled area and of the z-score-scaled y-size.

After learning the marker threshold on the P1-partition1, we validated its performance on three independent datasets: P1-partition2, P2 and P3. The rationale is to simulate a real scenario where the cell morpho-rheological features of three new patients (which we called P1-partion2, P2 and P3 and were never used during learning of the marker threshold) were analyzed with our marker. By applying the marker threshold, we computed for each of these potential patients the ability of our marker to predict the true label information (fluorescent probe labels are regarded as ground-truth in this study). In this particular validation, conceptually it does not make sense in our opinion to make a cross-validation, because we are evaluating a real scenario where three patients are going to the doctor and we compute for each of them the performance of our marker in comparison to ground-truth fluorescent probe. The result of this emulation of a realistic clinical estimation are provided in Table III and Supplementary Figure 2, where we display all the main statistics for evaluation of the classification of the cell types (mRBCs vs. RETs) of the three potential patients. However, since it could be also interesting to assess the performance of the investigated markers with 10-fold cross-validation on the 3 validation (patients) independent datasets, these results are provided in Suppl. Table I represented with the average performance on the 10 folds. We found that the overall accuracy on the three datasets is at the level of 0.74 and the overall AUC-ROC is around 0.70 (for P2, it reaches 0.76) with both patient validation and cross validation (Table III and Suppl. Table I). In general, AUC<0.6 is regarded as poor, while it is considered as acceptable if 0.7<AUC<0.8[25]. Therefore, the results here indicate that the designed combinatorial marker (based on area and y-size) together with the learned threshold can offer an acceptable classification performance on the independent validations, both internal and external. On the other hand, we could notice that the level of precision is very low (no higher than 0.05, please refer to Table III last row). This can be seen from the fact that, although the designed marker (and the respective threshold) can achieve an acceptable performance in correctly detecting RETs, on the other hand it makes a relevant false positive error by wrongly classifying a portion of mRBCs as RETs. This portion of wrongly classified mRBCs (which generate false positives) is small in comparison to the total amount of mRBCs, hence the overall specificity is around 0.74 (Table III), which is a relatively good value. However, since the dataset is unbalanced and the fraction of RETs is significantly smaller than mRBCs, even a small fraction of wrongly assigned mRBCs — since it is much larger than the total RETs – can cause a significant drop in precision. In order to demonstrate that the low precision is only due to the 'over-representation' of mRBCs and that the marker and threshold inferred are valid, we repeated the same validation analysis done in Table III considering mRBCs sampled at random in an equal amount to RETs (see Methods for details). The results reported in Figure 3 demonstrate that if, in the independent validation phase, we reduce the naturally occurring over-representation of mRBCs — using a procedure that is not biased, since it exploits a class-balance procedure based on random uniform mRBC sampling — then the level of precision increases drastically. Indeed, PC-corr marker increases precision from less than 0.05 (Table III, last line) to more than 0.82 (Figure 3, first bar in each plot). This shows that the low levels of precision do not originate from a learning error of the combinatorial marker and threshold, but from the over-representation (which can be interpreted as a sort of 'oversampling') of mRBCs in comparison to RETs. In practice, the low precision is generated by the fact that mRBCs are more abundantly represented in the dataset than RETs. This implies that, although the threshold is correct, the amount of mRBCs that pass the threshold assuming values similar to RETs is minor in comparison to the total amount of mRBCs. But it is still remarkable in comparison to the few total RETs present in the dataset. Taken together these results suggest that we mainly demonstrate the validity of the proposed unsupervised analysis pipeline as a proof of concept. For real-world application, higher level of precision is required, and the problem of unbalanced cell's cohorts should be adequately addressed in future studies.

**Table III**
**Validation Of The Proposed Morpho-Rheological Marker And Its Threshold On Three Independent Datasets**

| Performance | Internal validation based on P1-partion2 | External validation based on P2 | External validation based on P3 |
|---|---|---|---|
| Accuracy | 0.74 | 0.74 | 0.74 |
| Sensitivity | 0.65 | 0.77 | 0.63 |
| Specificity | 0.75 | 0.74 | 0.74 |
| AUC-ROC | 0.70 | 0.76 | 0.69 |
| Precision | 0.05 | 0.03 | 0.02 |

Finally, since the method used as reference for ab-initio labelling of the cells is based on fluorescence, we cannot assert that the mRBCs that pass the threshold assuming values of the proposed morpho-rheological marker similar to RETs are in general incorrectly assigned. In fact, to be more correct, we can only assert that there is a disagreement between our morpho-rheological marker assignment and the fluorescent assignment. Therefore, we can speculate that these mRBC cells, which are RET-like according to our morpho-rheological marker and not-RET-like according to fluorescence, should be investigated with more attention in future studies, because they might hide a cell sub-population in a 'gray area' that lies between mature red blood cells (mRBCs) and reticulocytes (RETs), which are immature red blood cells. A dichotomic separation between mRBCs and RETs might be over-simplistic, and a more truthful cell-phenotype landscape might consist of a fuzzy scenario populated also by intermediate and transition states. In fact, modern blood counters do distinguish different subpopulations of reticulocytes by their level of fluorescence. However, for the given measurements with the given gates, total reticulocyte numbers are in agreement with standard blood count performed at the university hospital.

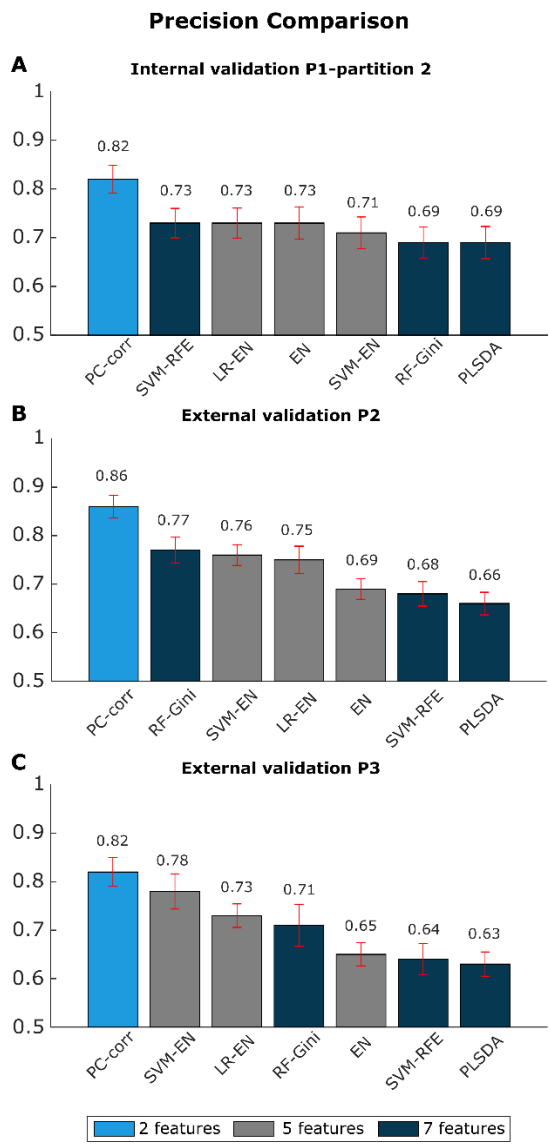

**Figure 3.** Mean precision performance of evaluated methods for an independent validation class-balanced scenario using 10,000 permutations of random uniform mRBC sampling. The bar color is associated with the number of features used to train the respective model. Light blue used two features; gray used five features; blue used seven features. The red whiskers report the standard deviation. A) Performance in internal validation P1-partition 2 data. B) Performance in external validation P2 data. C) Performance in external validation P3 data.

*F. Performance comparison with supervised approach*

The motivation for this unsupervised approach is the fact that the data are highly unbalanced, with 10,763 mRBCs (negative class) versus 257 RETs (positive class) for the machine learning model generation. It is known that regular supervised methods do not work well in these scenarios because they tend to predict new samples as the majority class in the training set, since these models try to optimize by accuracy. Moreover, Smialowki and colleagues demonstrated that PCA-based feature selection was more robust and less prone to overfitting in their study [17]. However, in order to quantitatively evaluate the extent to which our approach (which is based on unsupervised learning in the first part) offers better results, we compared it with the following supervised machine learning methods (which were trained according to the class-balance strategy reported in the methods section H): Support Vector Machine (SVM), Linear Regression (LR), Random Forest (RF) and Partial Least Squares Discriminant Analysis (PLSDA) using different feature selection strategies such as Gini index (for RF), Recursive Feature Elimination (for SVM) and elastic net (for LR and SVM). In addition, Elastic Net was also used as an all-in-one feature selection and classification method. Thus, we compared in total 6 supervised models versus our approach.

The results which show the performance in validation of the other MLs on three independent datasets is reported in Suppl. Table II, which should be compared with Table III for PC-corr. In order to simplify this investigation, we created a Table IV that summarizes the contrast between PC-corr combinatorial marker (based only on the average of two measures) and the other MLs combinatorial markers (based in general on models that adopt from 5 to 7 features, according to supervised feature selection). The comparison consists in counting how many times PC-corr performs better than the other methods considering 5 different evaluation measures in 3 independent validation sets (15 evaluations in total). Remarkably, from Table IV emerges that PC-corr using only two features provided a performance comparable and often higher (in 5 of the 6 comparisons) than more complicated models based on different machine learning rationales which use 5 to 7 features. In addition, in Figure 3 we compare the precision increase of our method versus the increase of the other MLs, when the validation on the three independent datasets is balanced by random uniform mRBC sampling. The extensive comparison provided here demonstrates that our proposed method is well performing in this challenging classification task also in comparison to state-of-the-art supervised methods. Taken together, the advantage of PC-corr is twofold: (i) it offers, using only two features, comparable performance to state-of-the-art methods that need from 5 to 7 features; (ii) it provides remarkable higher precision performance in comparison to state-of-the-art methods (Figure 3) when the datasets are balanced by random uniform mRBC sampling.

**Table IV**
**Comparison Of Performance Between PC-corr Based Marker Vs Other Machine Learning Based Markers On Three Independent Datasets.** The first column indicates the ML-based combinatorial markers based on the number of features indicated in brackets. The second column indicates the number of cases in which (comparing Table III of PC-corr validation with the respective tables of the ML-methods reported in Suppl. Table II) ML-methods perform better than PC-corr. The third column indicates the number of cases in which PC-corr (whose combinatorial marker is based on two features, which is the value reported in brackets near PC-corr name) performs better than other MLs. The fourth column reports the number of cases that are tied. Bold characters emphasize the number of times that a method performs better than PC-corr or vice versa. Remarkably, PC-corr performs better in 5 of the 6 comparisons.

| Methods | # Higher Values | PC-corr-based (2) | Tied |
|---|---|---|---|
| SVM EN (7) | **6** | 5 | 4 |
| LR-EN (7) | 6 | **7** | 2 |
| RF-Gini (5) | 5 | **8** | 2 |
| SVM RFE (5) | 4 | **10** | 1 |
| EN (7) | 4 | **10** | 1 |
| PLSDA (5) | 3 | **12** | 0 |

IV. DISCUSSION

RT-FDC is a powerful microfluidic technique[9] for the morpho-rheological characterization of cells and its correlation with conventional fluorescence-based analysis. Its high-throughput capability allows for efficient measurements also in cases with scarce populations such as reticulocytes. In this study, we demonstrated that the morpho-rheological features obtained with RT-FDC can be exploited to develop promising label-free combinatorial markers for cell biology research. First, we proposed a general computational and unsupervised machine learning framework for the design of combinatorial morpho-rheological markers and the marker-threshold definition. Our aim was to explore the potential and limitations of using a basic unsupervised and linear approach. We were interested in defining a baseline that could suggest what is possible to achieve using a simple and easily interpretable combination of morpho-rheological features to design a combinatorial marker for direct cell classification.

Our computational framework was proven in multiple independent validations to be able to provide acceptable performance when applied to a challenging (unbalanced-dataset) classification task such as the one to classify mRBCs vs RETs. This result is very promising and we hope that future studies investigate and address the current limitation of the methodology. Despite the acceptable level of classification, RETs are detected with low precision which, in combination with their naturally low prevalence, is problematic for some real-world applications. A significant number of cells, classified as mRBCs by their lack of RNA content staining, are falsely assigned to the RET population. This disagreement between the morpho-rheological and the fluorescence-based cell assignment, needs further investigation because it might indicate an error of detection. More interestingly, it could also suggest the presence of hidden, uncategorized sub-populations of cells.

Although the methods and findings provided here need further investigation to be used in clinical applications, they demonstrate the predictive potential of morpho-rheological phenotyping for computational-driven cell characterization/classification. Therefore, we expect that this study could contribute to the definition of new standards of analysis in precision and systems biomedicine.

Since this study was intended to evaluate how to implement a basic unsupervised and linear approach to discriminate mature RBCs and reticulocytes in the blood of an individual by using morpho-rheological phenotype data obtained from RT-FDC, future studies might go beyond this and investigate: i) more advanced approaches based on nonlinear machine learning directly on morpho-rheological data; and ii) deep learning techniques applied directly on the image samples, which could improve the performance of the classification without the pre-processing step to extract morpho-rheological features from the images.

## V. CONCLUSION

We propose an interdisciplinary study that deals with cell labeling from a different perspective by combining biophysics with machine intelligence tools. We started with a newly developed high-throughput single cell mechanics measurement technology, named real-time deformability and fluorescence cytometry (RT-FDC), and then we applied unsupervised machine learning to predict the labels of single cells, in particular we consider the task to discriminate and classify mature red blood cells against reticulocytes, which are immature red blood cell. We focus our study on this specific task because the investigation of reticulocytes (their percentage and cellular features) in the blood is important to quantitatively evaluate conditions that affect RBCs, such as anemia or bone marrow disorders. Our results suggest that the proposed machine intelligence data-driven methodology can provide promising results for the morpho-rheological-based prediction of red blood cells, therefore it can point out a new complementary direction to fluorescent cell labeling.

## AUTHOR CONTRIBUTION

CVC and JG conceived the study. CVC and YG designed the computational procedure. PR and JG devised and carried out the RT-FDC experiments. NT was responsible for acquisition of the blood samples. YG performed the computational analysis with CVC help. YG realized the MATLAB code and SC tested for quality control. CD realized the MATLAB code and performed the computational analysis for the supervised machine learning part. YG and CVC designed the figures and tables. YG and CVC wrote the article with inputs and corrections from all the other authors. YG and SC realized the figures under the CVC guidance. JG led, directed and supervised the study for the cell mechanics part. CVC led, directed and supervised the study for the computational part.

## ADDITIONAL INFORMATION

Competing financial interests

PR is holding shares of Zellmechanik Dresden GmbH, that sells devices based on RT-DC.

## ACKNOWLEDGMENT

We thank Alejandro Rivera Prieto and Salvatore Girardo, head of the Microstructure Facility of the Center for Molecular and Cellular Bioengineering (CMCB) at the Technische Universität Dresden (in part funded by the Sächsisches Ministerium für Wissenschaft und Kunst and the European Fund for Regional Development) for help with microfluidic chip production.

*Funding:* The work in C.V.C. research group was mainly supported by the independent group leader starting grant of the BIOTEC at Technische Universität Dresden (TUD) and by the Klaus Tschira Stiftung (KTS) gGmbH, Germany, grant name: Lipidom Signaturen fuer Alzheimer und Multiple Sklerose (Code: 00.285.2016). S.C. wrap-up postdoc period was supported by the Dresden International Graduate School for Biomedicine and Bioengineering (DIGS-BB), granted by the Deutsche Forschungsgemein-schaft (DFG) in the context of the Excellence Initiative. C. D. is supported by the Research Grant – Doctoral Programs in Germany from the Deutscher Akademischer Austauschdienst (DAAD), Promotion program Nr: 57299294. We acknowledge support by the German Research Foundation and the Open Access Publication Funds of the TU Dresden. Financial support from the Alexander-von-Humboldt Stiftung (Humboldt-Professorship to J.G.) and the DFG KFO249 (GU 612/2-2 grant to J.G.) is gratefully acknowledged.

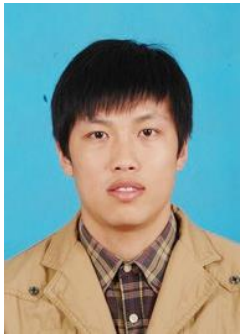

**Yan Ge** received the B.Sc. degree in biotechnology from the Southwest University of Science and Technology, China, in 2007, and the PhD degree in China Agricultural University in 2012. He was postdoc at the Biotechnology Center (BIOTEC) of Technische Universität Dresden from 2015 to 2018, under the joint supervision of Dr. Carlo Vittorio Cannistraci and Dr. Jochen Gück. Now he works as postdoc in the Medical Theoratcial Center (MTZ) of Universitätsklinikum Carl Gustav Carus Dresden. His research interests include single cell omics, machine learning and computational immunology.

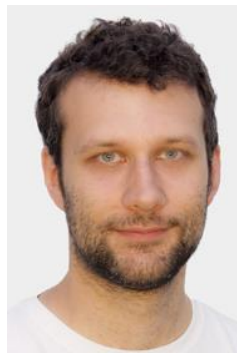

**Philipp Rosendahl**, received his Diploma and Ph.D. degree in Physics from Technische Universität Dresden, Germany in 2013 and 2018, respectively.
At the Biotechnology Center (BIOTEC) he was developing the methods RT-(F)DC and applying them to biological questions under supervision of Prof. Jochen Guck (Nature methods 2015 & 2018). He is now Head of Product Development and co-founder of the company *Zellmechanik Dresden GmbH*, which is making RT-DC commercially available. His research focusses on the technical development of microfluidic real-time systems for cytometry applications.
Dr. Rosendahl was awarded with the *Georg-Helm-Prize* for his Diploma thesis in 2013, the *Klaus-Goerttler-Prize* for his dissertation and the *IBA Best-Paper award* for the RT-FDC method publication in 2018.

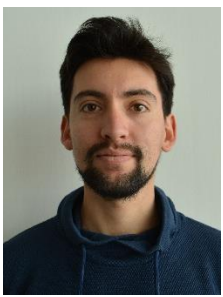

**Claudio Durán** received the Engineer diploma degree in bioinformatics from the University of Talca, Chile, in 2017. He is currently a PhD student in the Biomedical Cybernetics lab led by Dr. Carlo Vittorio Cannistraci in Technische Universität Dresden. His research interests include machine learning, network science and systems biomedicine.

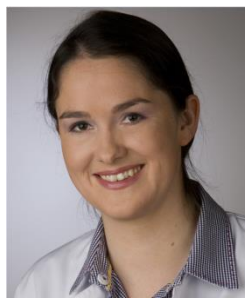

**Nicole Toepfner** received her MD degree and approbation at the Heinrich-Heine University Düsseldorf, Germany and specialized in Pediatrics at the University Medical Centers of Freiburg (2008-2012) and Dresden (2012-2014), Germany. She was a scholarship student of the DFG graduate school 320, an IFMSA scholarship holder (2006) and a DFG Gerok fellow (2014). Her work on streptococcal infections was awarded by the Young Investigator Award of the German Society for Pediatric Infectious Disease (2013). Before her specialist training in Pediatric Hematology and Oncology, Nicole Toepfner was a postdoc in the group of Prof. Jochen Guck, BIOTEC Dresden, Germany and of Prof. Edwin R. Chilvers, University of Cambridge, UK (2014-2016). She developed a label-free human blood cell analysis capable of detecting functional immune cell changes (e.g. of post primed neutrophils) in infection and inflammation (Elife 2018, Frontiers in Immunology 2018, Journal of Leukocyte Biology 2019).

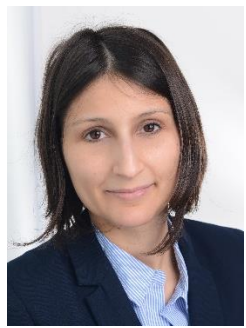

**Sara Ciucci** received the B.Sc. degree in Mathematics from the University of Padova, Italy, in 2010, the M.Sc. degree in Mathematics from the University of Trento, Italy, in 2014, and the PhD degree in Phyisics from Technische Universität Dresden, Germany, in 2018, at the Biomedical Cybernetics lab under the supervision of Dr. Carlo Vittorio Cannistraci. She is currently a postdoctoral researcher at the Biotechnology Center (BIOTEC) of Technische Universität Dresden in the Biomedical Cybernetics lab. Her research interests include machine learning, network science and systems biomedicine.

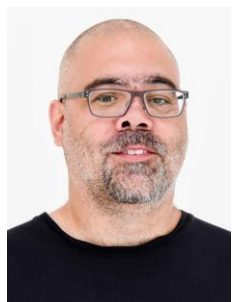

**Jochen Guck** is a cell biophysicist, born in Germany in 1973. He obtained his PhD in Physics from the University of Texas at Austin in 2001. After being a group leader at the University of Leipzig, he moved to the Cavendish Laboratory at Cambridge University as a Lecturer in 2007 and was promoted to Reader in 2009. In 2012 he became Professor of Cellular Machines at the Biotechnology Center of the Technische Universität Dresden. As of October 2019 he is now Director at the Max Planck Institute for the Science of Light and the Max-Planck-Zentrum für Physik und Medizin in Erlangen, Germany. His research centers on exploring the physical properties of biological cells and tissues and their importance for their function and behavior. He also develops novel photonic, microfluidic and scanning-force probe techniques for the study of these optical and mechanical properties. The ultimate goal is utilizing this insight for novel diagnostic and therapeutic approaches. He has authored over 100 peer-reviewed publications and four patents. His work has been recognized by several awards, amongst them the Cozzarelli Award in 2008, the Paterson Prize in 2011 and an Alexander-von-Humboldt Professorship in 2012.

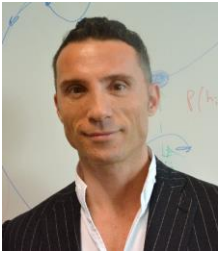

**Carlo Vittorio Cannistraci** is a theoretical engineer and was born in Milazzo, Sicily, Italy in 1976. He received the M.S. degree in Biomedical Engineering from the Polytechnic of Milano, Italy, in 2005 and the Ph.D. degree in Biomedical Engineering from the Inter-polytechnic School of Doctorate, Italy, in 2010. From 2009 to 2010, he was visiting scholar in the Integrative Systems Biology lab of Dr. Trey Ideker at the University California San Diego (UCSD), CA, USA. From 2010 to 2013, he was postdoc and then research scientist in machine intelligence and complex network science for personalized biomedicine at the King Abdullah University of Science and Technology (KAUST), Saudi Arabia. Since 2014, he has been Independent Group Leader and Head of the Biomedical Cybernetics lab at the Biotechnological Center (BIOTEC) of the TU-Dresden, Germany. He is also affiliated with the MPI Center for Systems Biology Dresden and with the Tsinghua Laboratory of Brain and Intelligence. He is author of three book chapters and more than 40 articles. His research interests include subjects at the interface between physics of complex systems, complex networks and machine learning theory, with particular interest for applications in biomedicine and neuroscience.

Dr. Cannistraci is member of the Network Science Society, member of the International Society in Computational Biology, member of the American Heart Association, member of the Functional Annotation of the Mammalian Genome Consortium. He is an Editor for the mathematical physics board of the journal Scientific Reports edited by Nature and of PLOS ONE. *Nature Biotechnology* selected his article (*Cell* 2010) on machine learning in developmental biology to be nominated in the list of 2010 notable breakthroughs in computational biology. *Circulation Research* featured his work (*Circulation Research* 2012) on leveraging a cardiovascular systems biology strategy to predict future outcomes in heart attacks, commenting: "a space-aged evaluation using computational biology". In 2017, Springer-Nature scientific blog highlighted with an interview to Dr. Cannistraci his recent study on "How the brain handles pain through the lens of network science". In 2018, the American Heart Association covered on its website Dr. Cannistraci's chronobiology discovery on how the sunshine affects the risk and time onset of heart attack. The TU-Dresden honoured Dr. Cannistraci of the *Young Investigator Award 2016 in Physics* for his recent work on the local-community-paradigm theory and link prediction in monopartite and bipartite complex networks.